\documentclass[reprint,amsmath,amssymb,aps]{revtex4}

\usepackage{hyperref}
\newcommand{\pfrac}[2]{\frac{\partial{#1}}{\partial{#2}}}

\renewcommand{\vec}[1]{\mathbf{#1}}
\newcommand{\afffias}{Frankfurt Institute for Advanced Studies (FIAS), Ruth-Moufang-Strasse~1, 60438 Frankfurt am Main, Germany}
\newcommand{\affgu}{Goethe University, Max-von-Laue-Strasse 1, 60438 Frankfurt am Main}

\begin{document}

\preprint{APS/123-QED}

\title{Covariant Canonical Quantization}

\author{P. Liebrich}
\affiliation{\afffias}
\affiliation{\affgu}

\date{\today}

\begin{abstract}
A formulation of Covariant Canonical Quantization is discussed, which works on an extended Hilbert space and reduces to conventional canonical quantization when constraining to the solution of the field equation \emph{a priori}. From the formal point of view it may be seen as a formalism between the canonical operator and the functional integral approach. A covariant number operator and two symmetric vacua are constructed. By that means, certain well-known quantities like the LSZ formula are rederived via a projection limit. The time-ordering operator can be replaced by taking into account the mirrored vacuum as well. Then the quantum field theoretical divergences like the vacuum energy arise \emph{a posteriori} when a spacetime split is performed. The role of the vacuum energy in different contexts is then discussed in general.
\end{abstract}

\maketitle

\section{Introduction}
In the path integral formalism the classical path is only the dominant mode and so it seems unsatisfying to restrict the quantum system to this mode right from the beginning. In gauge theories the ``covariant approach'' is an established method, but still only covers the gauge freedom. This formulation goes one step further and already impacts simple matter fields. Because the integration over all field configurations in the functional integral formalism is horrible from a mathematical point of view, the aim of this article is to construct a manifestly covariant operator theory and show how the canonical formulations are recovered when performing subtle reductions. Thereby, important physical properties can be uncovered which are otherwise hidden. The \textbf{LSZ formula} is rederived as an on-shell limit, and the \textbf{zero-point} or \textbf{vacuum energy} turns out to be ``hidden'' in a covariant number operator and becomes infinite as soon as a spacetime split is performed.

The word ``covariant canonical quantization'' has also been used in other contexts. The book \cite{NakanishiOjima} contains an extensive collection of the operator formalism in quantum field theory with a focus on gauge theory; the method studied here is still something else. In \cite{Peierls} a bracket is constructed that is built directly from the action and without reference to a certain phase space; here we go rather the opposite way. In \cite{HippelWohlfarth} this phrase is used to name an alternative quantization rule with which the Dirac equation is derived as a byproduct, but which deviates from conventional quantization for parameter spaces with dimension $d > 1$. One should also mention \cite{Helein} where a similar result emerged that the vacuum energies were first of all vanishing and only in the end acquired the usual infinite values by means of a metaplectic correction. This already comes close to the ideas employed here.

In the following, the procedure of \textbf{Covariant Canonical Quantization} is outlined. The limits to the conventional theory are shown and some implications are discussed. The reader should be aware that several quantities are denoted with the same letters like in standard canonical quantization, they are indeed analogous, but they are not necessarily equal.

\section{Covariant Canonical Quantization}
\subsection{Quantization postulates}
In the simplest version of canonical quantization, a real scalar field $\phi$ is promoted to an operator $\hat{\phi}$ and Fourier expanded as
\begin{equation}\label{FourierSkalar}
\hat{\phi}(x) = \int \frac{1}{(2\pi)^4} \hat{\tilde{\phi}}(k) e^{-ik_\mu x^\mu} d^4k.
\end{equation}
(Natural units $\hbar = c = 1$ are employed and the metric convention is $(+,-,-,-)$.) The usual formulations, see e.~g. \cite{Fock, GreinerReinhardt, Srednicki, PeskinSchroeder}, now insert this into the Klein-Gordon equation and, therefore, receive a split of the coefficients in terms of ``creation'' and ``annihilation'' operators. This means that they work on the space of \emph{solutions} of the field equation, which is unsatisfying as this incorporates \emph{classical} information into the \emph{quantum} field right from the beginning. Of course, the conventional theory is equally valid, but some peculiarities like sudden infinities lack a proper limit, which would justify the renormalization limits. The approach here uses an unrestricted Hilbert space and in the end the off-shell modes will be forced to vanish, much like the covariant treatment of gauge theories. In those cases an additional layer is opened up, namely
\begin{equation}\label{Hilbert}
\mathfrak{H}_\text{real} \subset \mathfrak{H}_\text{phys} \subset \mathfrak{H}_\text{total}.
\end{equation}
In comparison with the functional integral formalism, the mathematical complexity of integration over all fields can be omitted. Instead, the information is fully contained in the commutators of the operators. They describe all particles, including the ``virtual'' ones, but in the end only the ``real'' ones will contribute to the observables. It is already known from quantum \emph{mechanics} that for relativistic systems one has to embed the motion into an extended phase space to get a well-defined global formulation of the problem and then constrain to the mass-shell and this leads to the correct Klein-Gordon instead of the Schr\"odinger equation \cite{ExtPoint,ExtPointGeom}.

The common starting point of quantization is to promote the Poisson bracket to the equal-time commutation relation:
\begin{equation}
\left[\hat{\phi}(\vec{x},t), \hat{\pi}^\mu(\vec{y},t)\right] \overset{!}{=} i \delta_0^\mu \delta^{(3)}(\vec{x}-\vec{y}).
\end{equation}
In this formula all conjugate momenta shall be represented, while usually only the 0-component is paid attention to, but it can be calculated that the other commutators have to vanish anyway. The more important point here is the commutator for general spacetime separations. Together with the Klein-Gordon Lagrangian one can calculate that the respective commutator for a free scalar field is described by the \textbf{Pauli-Jordan function}\index{Pauli-Jordan function}
\begin{equation}
\Delta(x-y) := -\int \frac{1}{(2\pi)^3 E_\vec{k}} e^{-ik_i (x^i - y^i)} \sin(E_\vec{k} (x^0 - y^0)) d^3\vec{k}.
\end{equation}
With some identities for the $\delta$ distribution it can be rewritten:
\begin{equation}
\Delta(x-y) = i \int \frac{2}{(2\pi)^3} \delta(k_\mu k^\mu - m^2) \operatorname{sign}(k^0) e^{-ik_\nu (x^\nu - y^\nu)} d^4k.
\end{equation}

The point of departure from ``standard'' quantization is to take the commutators as the fundamental postulates. So although the Klein-Gordon Lagrangian may be used to derive the above relation, only that commutator postulate will be used for now, creating the extra layer in (\ref{Hilbert}). Therefore, the field will only be called a ``scalar field'' in the following. This yields for the commutator of the Fourier modes:
\begin{equation}\label{KommFourierSkalar}
\left[\hat{\tilde{\phi}}(k), \hat{\tilde{\phi}}(k')\right] \overset{!}{=} (2\pi)^5 \delta^{(4)}(k + k') \delta(k_\mu k^\mu - m^2) \operatorname{sign}(k^0).
\end{equation}
The prefactor is a normalization convention. The term $\operatorname{sign}(k^0)$ distinguishes the time direction, thus making the commutator well-defined, as it must be anti-symmetric. The presence of the 0-component of the 4-momentum in $\operatorname{sign}(k^0)$ restricts the invariance of (\ref{KommFourierSkalar}) under Lorentz transformations to \emph{orthochronous} ones.

\subsection{Covariant particle creation and annihilation}
For a meaningful particle interpretation it is useful to split the Fourier expansion (\ref{FourierSkalar}) as follows:
\begin{equation}\label{FourierSkalarSplit}
\begin{aligned}
\hat{\phi}(x) &= \int \frac{1}{(2\pi)^4} \hat{\tilde{\phi}}(k) e^{-ik_\mu x^\mu} d^4k \\
&= \int \frac{1}{(2\pi)^4} \left(\frac{1}{2} \hat{\tilde{\phi}}(k) e^{-ik_\mu x^\mu} + \frac{1}{2} \hat{\tilde{\phi}}(-k) e^{ik_\mu x^\mu}\right) d^4k \\
&= \int \frac{1}{(2\pi)^4} \left(\hat{a}(k) e^{-ik_\mu x^\mu} + \hat{a}^\dagger(k) e^{ik_\mu x^\mu}\right) d^4k,
\end{aligned}
\end{equation}
where we defined the \textbf{covariant creation operator}
\begin{equation}
\hat{a}^\dagger(k) := \frac{1}{2} \hat{\tilde{\phi}}(-k),
\end{equation}
and the \textbf{covariant annihilation operator}
\begin{equation}
\hat{a}(k) := \frac{1}{2} \hat{\tilde{\phi}}(k),
\end{equation}
including on-shell and off-shell particles. It can readily be seen that they are not independent of each other. Their commutator
\begin{equation}\label{OpKommSkalar}
[\hat{a}(k), \hat{a}^\dagger(k')] = \frac{(2\pi)^5}{4} \delta^{(4)}(k - k') \delta(k_\mu k^\mu - m^2) \operatorname{sign}(k^0)
\end{equation}
has the expected sign in the energy-momentum-$\delta$, justifying the shape of the commutator (\ref{KommFourierSkalar}).

Furthermore, define the \textbf{covariant number operator}
\begin{equation}
\hat{N}(k) := \frac{4}{(2\pi)^5} \hat{a}^\dagger(k)\,\hat{a}(k).
\end{equation}
This operator generates a Fock space as the product of eigenstates $|n_k\rangle$ of $\hat{N}(k)$. We set
\begin{equation}\label{ZahlZust}
\hat{N}(k) |n_k\rangle =: (\delta^{(4)}(0) - \delta^{(4)}(2k))\,\delta(k_\mu k^\mu - m^2) \operatorname{sign}(k^0) n_k |n_k\rangle.
\end{equation}
Of course, it now needs to be shown that the operators $\hat{a}^\dagger$ and $\hat{a}$ do really act as ``creation'' and ``annihilation'' operators. It is already clear from the definition that they are a mix of the respective on-shell operators and so $\hat{N}(k)$ has contributions of two ``creation'' resp. two ``annihilation'' operators depending on the 4-momentum, but this does not yet mean much for its spectrum. It is straightforward to calculate
\begin{equation}\label{ErzSkalar}
\begin{aligned}
\hat{N}(k) \hat{a}^\dagger(k) |n_k\rangle &= \hat{a}^\dagger(k) \hat{N}(k) |n_k\rangle + [\hat{N}(k), \hat{a}^\dagger(k)] |n_k\rangle \\
&= (\delta^{(4)}(0) - \delta^{(4)}(2k)) \delta(k_\mu k^\mu - m^2) \operatorname{sign}(k^0) (n_k + 1) \hat{a}^\dagger(k) |n_k\rangle
\end{aligned}
\end{equation}
and
\begin{equation}\label{VernSkalar}
\begin{aligned}
\hat{N}(k) \hat{a}(k) |n_k\rangle &= \hat{a}(k) \hat{N}(k) |n_k\rangle + [\hat{N}(k), \hat{a}(k)] |n_k\rangle \\
&= (\delta^{(4)}(0) - \delta^{(4)}(2k)) \delta(k_\mu k^\mu - m^2) \operatorname{sign}(k^0) (n_k - 1) \hat{a}(k) |n_k\rangle.
\end{aligned}
\end{equation}
In other words, $n_k$ allows in fact for an interpretation as ``particle number'' that counts particles with 4-momentum $k$ on mass-shell. Creation operators effectively become annihilation operators and vice versa for negative energy states. The Dirac $\delta$'s in front are only a formality; in discretized spacetime they are 1 at 0 and 0 otherwise. But for massless particles at rest ($k = 0$) they tell us that nothing happens: the particle number loses its meaning, which is intuitively clear as there is nothing but a non-existing particle.

\subsection{Vacuum and energy split}
A special feature of these operators is their complete symmetry. But this is incompatible with $$|1 \cdot k\rangle \propto \hat{a}^\dagger(k) |0\rangle = \hat{a}(-k) |0\rangle = 0.$$ So the symmetry has to be broken at this point. This suggests that the (real) \textbf{vacuum} $|0\rangle$ has to be defined as the state where the application of the annihilation operator \emph{with positive energy} leads to a 0, which is equivalent to the action of the creation operator with negative energy:
\begin{equation}\label{Vakuum}
\Theta(k^0) \hat{a}(k) |0\rangle := 0 \iff \Theta(-k^0) \hat{a}^\dagger(k) |0\rangle := 0 \quad \forall k.
\end{equation}
Thereby, the standard vacuum will be recovered.

But it can also be seen that mathematically there is another Fock space that is built by the restriction on negative energies with an ``anti-vacuum'' $|0_-\rangle$. This is a general feature of formulations that work on an extended phase space and project down for the physical solution. Remember e.~g. the simple special relativistic energy-momentum relation $$E^2 = p^2 + m^2.$$ It has a solution with positive and negative energy, but the states realized in nature only carry positive energy. A reason for a sharp split is that the two mass-shells are disjoint, therefore a system in one shell will never reach the other shell. So it is necessary to pick one ladder. It can now easily be shown that the normalization factors for positive energy are real and those for negative energy are imaginary. This leads to the decision to take the positive vacuum state as the canonical one. However, if all physical processes were described with negative energy, then the situation would turn around: In that case the mirrored vacuum could and would have to be chosen. So this formulation allows for the flexibility that conventional canonical quantization does not possess.

Note that such a description is indeed physically allowed: Tachyons are conventionally only a mathematical construct. An observer sitting on the top of a mountain, looking into the night sky and spinning around, would experience the stars moving with speed faster than light. This is, of course, only a formal change of the reference system to calculate from. But it is equally valid, and so is a description with respect to the mirrored state space.

The same reasoning can be applied directly to the particle operators. The results from (\ref{ErzSkalar}) and (\ref{VernSkalar}) show that \emph{real} quantities should be defined as constrained on the upper mass-shell. Define $E_\vec{k} := \sqrt{\vec{k} \cdot \vec{k} + m^2}$. The \emph{real} creation operator thus is
\begin{equation}
\int \hat{a}^\dagger(k) \delta(k_\mu k^\mu - m^2) \Theta(k^0) dk^0 = \int \frac{\hat{a}^\dagger(k)}{2k^0} \delta(k^0 - E_\vec{k}) dk^0 = \frac{\hat{a}^\dagger(\vec{k},E_\vec{k})}{2E_\vec{k}}
\end{equation}
and the \emph{real} annihilation operator is
\begin{equation}
\int \hat{a}(k) \delta(k_\mu k^\mu - m^2) \Theta(k^0) dk^0 = \int \frac{\hat{a}(k)}{2k^0} \delta(k^0 - E_\vec{k}) dk^0 = \frac{\hat{a}(\vec{k},E_\vec{k})}{2E_\vec{k}}.
\end{equation}
These are the operators usually encountered in canonical quantization.

\subsection{Particle propagation}
In scattering calculations a special operation enters the description that explicitly breaks spacetime covariance. It is the time-ordering, denoted $\mathcal{T}(\cdot)$, to bring creation and annihilation operators on the right side of the expression. Now this is only necessary in conventional canonical quantization because there has already implicitly been made a symmetry break. In the formulation here it can alternatively be expressed with the two vacua, e.~g. the \textbf{Feynman propagator} reads
\begin{equation}\label{FeynmanProp}
\begin{aligned}
\Delta_F(x-y) &:= -i\langle0|\mathcal{T}(\hat{\phi}(x)\hat{\phi}(y))|0\rangle \\
&= -i\left((\Theta_+\langle0| + i\Theta_-\langle0_-|) \hat{\phi}(x)\hat{\phi}(y) (\Theta_+|0\rangle + i\Theta_-|0_-\rangle)\right) \\
&\left[= \frac{1}{(2\pi)^4} \int \frac{1}{k_\nu k^\nu - m^2 + i\epsilon} e^{-ik_\mu (x^\mu - y^\mu)} d^4k\right],
\end{aligned}
\end{equation}
where $\Theta_+ := \Theta(x^0 - y^0)$ and $\Theta_- := \Theta(y^0 - x^0)$, see appendix \ref{Feynman} for a derivation. It is an explicitly covariant quantity and here covariance never has to be given up, as opposed to the conventional formulation where this has to be compensated again by an artificial time-ordering. This expression seems more natural and, indeed, reduces the calculation effort for propagators by one half since one can skip the necessity to pay attention to both the energy and the particle operator type split.

One might also ask for a \textbf{covariant} $S$\textbf{-matrix}, in analogy to be defined by
\begin{equation}
S := \langle k_1,\ldots,k_m;\text{out}| l_1,\ldots,l_n;\text{in}\rangle
\end{equation}
with whole 4-momenta $k_i$ and $l_j$. These are the fundaments of scattering calculations as they describe events of in-particles with momentum $k$ and out-particles with momentum $l$. In appendix \ref{LSZFormel} the respective covariantized version of the \textbf{LSZ reduction formula} is derived as:
\begin{equation}
\begin{aligned}
\langle k;\text{out}| l;\text{in}\rangle &= \frac{-i}{4(2\pi)^6} \iint \left(\pfrac{}{x^\mu} \pfrac{}{x_\mu} - m^2\right) \left(\pfrac{}{y^\mu} \pfrac{}{y_\mu} - m^2\right) \\
&\quad \cdot \Delta_F(x-y) \delta(k_\nu k^\nu - m^2) \delta(l_\nu l^\nu - m^2) e^{ik_\xi x^\xi - il_\xi y^\xi} d^4x d^4y.
\end{aligned}
\end{equation}
It indeed reduces to the usual LSZ formula \cite{LSZ} on-shell. This result perfectly illustrates the distinction to the canonical formalism: There one effectively ``divides out'' the $\delta$-function. This shows why there are the renormalization issues; since the core formula is calculated by dividing by $\infty$, it comes at no surprise that this $\infty$ reappears in the numerator.
 
\section{Vacuum energy}
A special feature of the canonical operator theory as opposed to the functional integral formulation is that it allows to calculate ``conserved quantities''. Consider the \emph{canonical} \textbf{energy-momentum tensor} corresponding to the Klein-Gordon Lagrangian $\mathcal{L}_{KG}$:
\begin{equation}\label{EnImpTensor}
\theta_{\mu\nu} := \pfrac{\mathcal{L}_{KG}}{\pfrac{\phi}{x_\mu}} \pfrac{\phi}{x^\nu} - \eta_{\mu\nu} \mathcal{L}_{KG} = \pfrac{\phi}{x^\mu} \pfrac{\phi}{x^\nu} - \eta_{\mu\nu} \frac{1}{2} \left(\pfrac{\phi}{x^\xi} \pfrac{\phi}{x_\xi} - m^2 \phi^2\right).
\end{equation}
According to the Einstein field equations, the (expectation value of the) \emph{absolute} energy-momentum tensor contributes to the spacetime dynamics and so it is desirable to calculate its value. As one cannot find a direct expression for its elements in terms of a number operator, one is forced to integrate out the $x$-dependence and look at its representation in momentum space. The \textbf{instantaneous energy} of the field is defined as the space-integral of the 00-component of this tensor:
\begin{equation}
E(x^0) := \int \theta_{00}(x) d^3\vec{x}.
\end{equation}
In this covariant approach, the \textbf{total energy}, i.~e. its integral also over time, is the fundamental object. There seems to be no reason why time should play a preferred role already at this stage, particularly as the instantaneous energy is not the ultimate quantity in question. The corresponding quantum expression can be expanded in terms of the covariant creation and annihilation operators:
\begin{equation}\label{IntEnSkalar}
\begin{aligned}
\hat{E}_\text{tot} &:= \int \hat{E}(x^0) dx^0 \\
&= \frac{1}{2(2\pi)^8} \iiint (-k_0 k_0' + k^i k_i' + m^2) \hat{\tilde{\phi}}(k) \hat{\tilde{\phi}}(k') \\
&\quad\quad\quad\quad\quad\quad\quad \cdot e^{-i(k_\mu+k_\mu') x^\mu} d^4k d^4k' d^4x \\
&= \frac{2}{(2\pi)^4} \int (k_0^2 + E_\vec{k}^2) \hat{a}^\dagger(k) \hat{a}(k) d^4k \\
&= \pi \int (k_0^2 + E_\vec{k}^2) \hat{N}(k) d^4k.
\end{aligned}
\end{equation}
The quadratic dependence on the energy is a bit fallacious as the mass-shell, implicitly contained in $\hat{N}$ (compare eq. (\ref{ZahlZust})), actually reduces its order by 1. To see this explicitly take the trace. By means of the identities for the $\delta$ distribution this gives
\begin{equation}
\begin{aligned}
E_\text{tot} &:= \sum_n \int \langle n_k|\hat{E}_\text{tot}|n_k\rangle d^4k \\
&= \pi \sum_n \int (k_0^2 + E_\vec{k}^2) (\delta^{(4)}(0) - \delta^{(4)}(2k)) \delta(k_\mu k^\mu - m^2) \operatorname{sign}(k_0) n(k) d^4k \\
&= \pi \delta^{(4)}(0) \sum_n \int E_\vec{k} \left(n(\vec{k},E_\vec{k}) - n(-\vec{k},-E_\vec{k})\right) d^3\vec{k} \\
&=: 2\pi \delta^{(4)}(0) \sum_n \int E_\vec{k} n(\vec{k}) d^3\vec{k}.
\end{aligned}
\end{equation}
The last conversion shall be justified by the symmetry $\hat{a}^\dagger(k) = \hat{a}(-k)$ that creating a particle with negative energy is like annihilating a particle with positive energy. The symmetry could not hold directly on the state vector level due to some formal reasons, but can be recovered here. (Otherwise that would not alter the argumentation either.)

Does this mean that in CCQ there would be no vacuum energy? Actually, the \emph{vacuum} energy has not been calculated so far. When the covariant expression for the total energy (\ref{IntEnSkalar}) is applied to the canonical vacuum as defined in (\ref{Vakuum}), then the vacuum energy after all equals
\begin{equation}
\begin{aligned}
\langle0| \hat{E}_\text{tot} |0\rangle 
&= \frac{2}{(2\pi)^4} \int (k_0^2 + E_\vec{k}^2) \left(\Theta(k^0) \langle0| \hat{a}^\dagger(k) \hat{a}(k) |0\rangle\right. \\
&\quad\quad\quad \left.+ \Theta(-k^0) \langle0| \hat{a}(k) \hat{a}^\dagger(k) + [\hat{a}^\dagger(k), \hat{a}(k)] |0\rangle\right) d^4k \\
&= \pi \int (k_0^2 + E_\vec{k}^2) \Theta(-k^0) \delta(k_\mu k^\mu - m^2) \operatorname{sign}(-k^0) d^4k \\
&= 2\pi \int \frac{1}{2} E_\vec{k} d^3\vec{k}.
\end{aligned}
\end{equation}
This is exactly the conventional zero-point energy multiplied by a phase space factor of $2\pi$ that resembles the additional time integration. It comes from what had to be thrown away in definition (\ref{Vakuum}). There is in principle also another ``anti-Fock space'' that yields the same zero-point energy, that is why in the calculations before it did not appear. Considered separately, there are the usual divergences. So this split is a very crucial point and the rationale to arrive there already indicated that the vacuum is not ``empty'', but rather a state of ``broken particle-anti-particle symmetry''. In fact, it has the same origin as the explicit time-ordering causes the divergences. In contrast to the canonical approach, here one has to impose some ordering for \emph{receiving} the common zero-point energy density, rather than \emph{eliminating} it. This is a feature also seen in geometric quantization, where the final correction is called \textbf{metaplectic correction} \cite{Helein}. The representation with two vacua in the language of Covariant Canonical Quantization is physically more accessible and gives rationale to that operation.

\section{Discussion}
This paper outlined the method of Covariant Canonical Quantization with the key features not to restrict to the mass-shell in the expansion of creation and annihilation operators and to keep space and time on equal footing for as long as possible. In its mindset this procedure is similar to the functional integral formulation, but stays on the operator level. Standard canonical quantization is recovered by restricting to the on-shell Hilbert space a priori. In this formulation the vacuum energy is inherently incorporated in a ``covariant number operator'' and manifests by explicitly breaking the spacetime symmetry.

In view of this insight it may be worth to comment on the ``cosmological constant problem'' \cite{Weinberg:1988cp, Carroll, PaganiReuter}: There are two major ways how to fix the Einstein equation with quantum field theory
\begin{equation}
G_{\mu\nu} \approx 8 \pi G \hat{\theta}_{\mu\nu}:
\end{equation}
The left-hand side could be promoted to an operator as well and this is the task of quantum gravity. Another possibility is that only some expectation value of the right-hand side enters so that general relativity can stay classical. In that case there is the question which expectation value is to be taken. The canonical vacuum expectation value produces a cosmological constant $120$ orders too big to match observations. But the representation provided here suggests that it may not be the ultimate answer to this question and perhaps another expectation value has to be taken in that context that removes the divergence. In the literature there are statements that this vacuum energy is real, but might not contribute on astronomical scales. In a recent article something similar was argued via an initial data analysis as on the Planck scale there may be no arrow of time \cite{Carlip}. Another recent article uses a thermodynamic reasoning to argue that the vacuum energy must be zero when the vacuum equation of state shall be relativistically invariant \cite{Ryskin}.

The general problem in such derivations is that the source term is not the energy itself, but the energy \emph{density} and it is questionable whether simply calculating the total energy and then dividing it by the volume yields the correct answer. As discussed above, this quantity cannot be calculated directly, so the Dark Energy effect has to be derived in an indirect manner. Another derivation stems from the \textbf{Casimir force} \cite{Casimir}
\begin{equation}
F_C = \frac{\pi^2 A}{240d^4}
\end{equation}
as the force on the area $A$ between two conducting plates at distance $d$. This effect is generally seen to ``prove'' the physicality of the vacuum energy. In the formula it can already be seen that the force $F_C$ decreases with the distance $d$ and ultimately approaches 0 for infinite distance. If the Casimir effect shall account for Dark Energy, the borders indeed have to be set at infinity. That force results not only from the content inside the plates, but also from the force outside. When there is no ``outside'' any more, the pressure on the universe is simply non-existent. This matches with the observed value of the cosmological constant, which is very well around 0 compared to vacuum expectation calculations.

Fields with internal structure like Dirac or Maxwell fields were not covered here. However, similar features will emerge in a straightforward manner. Interacting theories can be discussed analogously by combining the single ingredients. Another interesting point arising in gauge theories is the gauge fixing and ghost machinery, and ultimately a quantum theory of gravity. This is to be discussed in the future.

\appendix
\section{Feynman propagator}\label{Feynman}
The Feynman propagator is the time-ordered version of the 2-point function which denotes the vacuum expectation value of 2 scalar fields with respect to the vacuum state $|0\rangle$. So this quantity can be calculated as
\begin{align*}
\langle0|\hat{\phi}(x)\hat{\phi}(y)|0\rangle &= \frac{1}{(2\pi)^8} \iint \langle0|\hat{\tilde{\phi}}(k) \hat{\tilde{\phi}}(k')|0\rangle e^{-ik_\mu x^\mu -ik_\mu' y^\mu} d^4k d^4k' \\
&= \frac{4}{(2\pi)^8} \iint \langle0|\hat{a}(k) \hat{a}^\dagger(k')|0\rangle e^{-ik_\mu x^\mu +ik_\mu' y^\mu} d^4k d^4k' \\
&= \frac{4}{(2\pi)^8} \iint \Theta(k_0) \Theta(k_0') \langle0|\hat{a}^\dagger(k') \hat{a}(k) + [\hat{a}(k), \hat{a}^\dagger(k')]|0\rangle e^{-ik_\mu x^\mu +ik_\mu' y^\mu} d^4k d^4k' \\
&= \frac{4}{(2\pi)^8} \iint \Theta(k_0) \Theta(k_0') \frac{(2\pi)^5}{4} \delta^{(4)}(k - k') \delta(k_\mu k^\mu - m^2) \operatorname{sign}(k^0) e^{-ik_\mu x^\mu +ik_\mu' y^\mu} d^4k d^4k' \\
&= \int \frac{1}{(2\pi)^3 2k^0} \Theta(k_0) (\delta(k^0 + E_\vec{k}) + \delta(k^0 - E_\vec{k})) e^{-ik_\mu (x^\mu - y^\mu)} d^4k \\
&= \int \frac{1}{(2\pi)^3 2E_\vec{k}} e^{-iE_\vec{k} (x^0 - y^0)} e^{-ik_i (x^i - y^i)} d^3\vec{k}. \\
\end{align*}
Now the same 2-point function with respect to the anti-vacuum state $$\Theta(-k^0) \hat{a}(k) |0_-\rangle := 0 \iff \Theta(k^0) \hat{a}^\dagger(k) |0_-\rangle := 0 \quad \forall k$$ is given by
\begin{align*}
\langle0_-|\hat{\phi}(x)\hat{\phi}(y)|0_-\rangle &= \frac{1}{(2\pi)^8} \iint \langle0_-|\hat{\tilde{\phi}}(k) \hat{\tilde{\phi}}(k')|0_-\rangle e^{-ik_\mu x^\mu -ik_\mu' y^\mu} d^4k d^4k' \\
&= \frac{4}{(2\pi)^8} \iint \langle0_-|\hat{a}(k) \hat{a}^\dagger(k')|0_-\rangle e^{-ik_\mu x^\mu +ik_\mu' y^\mu} d^4k d^4k' \\
&= \frac{4}{(2\pi)^8} \iint \Theta(-k_0) \Theta(-k_0') \langle0_-|\hat{a}^\dagger(k') \hat{a}(k) + [\hat{a}(k), \hat{a}^\dagger(k')]|0_-\rangle e^{-ik_\mu x^\mu +ik_\mu' y^\mu} d^4k d^4k' \\
&= \frac{4}{(2\pi)^8} \iint \Theta(-k_0) \Theta(-k_0') \frac{(2\pi)^5}{4} \delta^{(4)}(k - k') \delta(k_\mu k^\mu - m^2) \operatorname{sign}(k^0) e^{-ik_\mu x^\mu +ik_\mu' y^\mu} d^4k d^4k' \\
&= -\int \frac{1}{(2\pi)^3 2|k^0|} \Theta(-k_0) (\delta(k^0 + E_\vec{k}) + \delta(k^0 - E_\vec{k})) e^{-ik_\mu (x^\mu - y^\mu)} d^4k \\
&= -\int \frac{1}{(2\pi)^3 2E_\vec{k}} e^{iE_\vec{k} (x^0 - y^0)} e^{-ik_i (x^i - y^i)} d^3\vec{k}. \\
\end{align*}
For the calculation with both vacuum types it is important to note their disjoint nature: $$\langle0|\cdots|0_-\rangle = \langle0_-|\cdots|0\rangle = 0.$$ Now with $\Theta_+ := \Theta(x^0 - y^0)$ and $\Theta_- := \Theta(y^0 - x^0)$ and the residue theorem the Feynman propagator becomes
\begin{align*}
\Delta_F(x-y) &= \int \frac{1}{(2\pi)^4} \frac{e^{-ik_0(x^0-y^0)}}{(k^0 - \sqrt{E_\vec{k}^2 - i\epsilon}) (k^0 + \sqrt{E_\vec{k}^2 - i\epsilon})} e^{-ik_i (x^i - y^i)} d^4k \\
&= \int \frac{1}{(2\pi)^3 2E_\vec{k}} \left(\Theta(x_0 - y_0) e^{-iE_\vec{k}(x^0-y^0)} + \Theta(y_0 - x_0) e^{iE_\vec{k}(x^0-y^0)}\right) e^{-ik_i (x^i - y^i)} d^3\vec{k} \\
&= -i\Theta_+\langle0| \hat{\phi}(x)\hat{\phi}(y) |0\rangle + i\Theta_-\langle0_-| \hat{\phi}(x)\hat{\phi}(y) |0_-\rangle \\
&= -i\left((\Theta_+\langle0| + i\Theta_-\langle0_-|) \hat{\phi}(x)\hat{\phi}(y) (\Theta_+|0\rangle + i\Theta_-|0_-\rangle)\right),
\end{align*}

\section{Covariant LSZ formula}\label{LSZFormel}
Consider the simple covariant $S$\textbf{-matrix} of the form
\begin{equation}
S := \langle k;\text{out}| l;\text{in}\rangle
\end{equation}
and assume $k \neq l$. To define the in- and out-states in a way that the $S$-matrix reduces to the conventional one for on-shell momenta, one has to introduce a further ``time-like'' parameter. This is the price to pay when working with a Hilbert space defined on the total configuration space. In general, a time-independent operator is made time-dependent via the Heisenberg equation or a time-evolution operator of $$\hat{U}(t_0,t_1) = \mathcal{T}\left(\exp\left(-i\int_{t_0}^{t_1}\hat{H}(t)dt\right)\right).$$ In the covariant formalism the principle stays the same, but the required Hamiltonian is now the parametrized one like in \cite{ExtPoint}, which essentially represents the mass-shell condition. This leads to
\begin{equation}
\hat{a}_\text{out/in}^\dagger(k) := \lim_{s\to\pm\infty} \hat{a}^\dagger(k,s) := \lim_{s\to\pm\infty} \hat{a}^\dagger(k) e^{-i(k_\mu k^\mu - m^2)s}.
\end{equation}
A key relation to calculate is then the difference between the in- and out-creation operator:
\begin{equation}\label{InOut}
\begin{aligned}
\zeta^\dagger(k) := \hat{a}_\text{in}^\dagger(k) - \hat{a}_\text{out}^\dagger(k) &= -\int_{-\infty}^\infty \partial_s \hat{a}^\dagger(k,s) ds \\
&= -\int_{-\infty}^\infty \partial_s \left(\hat{a}^\dagger(k) e^{-i(k_\mu k^\mu - m^2)s}\right) ds \\
&= i(k_\mu k^\mu - m^2) \hat{a}^\dagger(k) 2\pi \delta(k_\nu k^\nu - m^2) \\
&= \frac{i}{2(2\pi)^3} \int (k_\mu k^\mu - m^2) \hat{\phi}(x) e^{-ik_\xi x^\xi} \delta(k_\nu k^\nu - m^2) d^4x \\
&= \frac{i}{2(2\pi)^3} \int \hat{\phi}(x) (\partial_\mu \partial^\mu - m^2) e^{-ik_\xi x^\xi} \delta(k_\nu k^\nu - m^2) d^4x \\
&= \frac{i}{2(2\pi)^3} \int e^{-ik_\xi x^\xi} (\partial_\mu \partial^\mu - m^2) \hat{\phi}(x) \delta(k_\nu k^\nu - m^2) d^4x.
\end{aligned}
\end{equation}
The last line used integration by parts; the surface term can be argued away like in the standard formulation.

To make the in- and out-modes physical, the limit $s \to t$ (parameter time to physical time) has to be taken. So what has to be calculated in the end can be reduced to (assume $k \neq l$)
\begin{equation}
\begin{aligned}
\langle k;\text{out}| l;\text{in}\rangle &= \langle 0| \hat{a}_\text{out}(k) \hat{a}_\text{in}^\dagger(l) |0\rangle \\
&= (\Theta_+\langle0| + i\Theta_-\langle0_-|) \hat{a}_\text{out}(k) \hat{a}_\text{in}^\dagger(l) (\Theta_+|0\rangle + i\Theta_-|0_-\rangle) \\
&= (\Theta_+\langle0| + i\Theta_-\langle0_-|) (\hat{a}_\text{in}(k) - \zeta(k)) (\hat{a}_\text{out}^\dagger(l) + \zeta^\dagger(l)) (\Theta_+|0\rangle + i\Theta_-|0_-\rangle) \\
&= \frac{-1}{4(2\pi)^6} \iint \left(\pfrac{}{x^\mu} \pfrac{}{x_\mu} - m^2\right) \left(\pfrac{}{y^\mu} \pfrac{}{y_\mu} - m^2\right) \\
&\quad \cdot (\Theta_+\langle0| + i\Theta_-\langle0_-|) \hat{\phi}(x) \hat{\phi}(y) (\Theta_+|0\rangle + i\Theta_-|0_-\rangle) \delta(k_\nu k^\nu - m^2) \delta(l_\nu l^\nu - m^2) e^{ik_\xi x^\xi - il_\xi y^\xi} d^4x d^4y \\
&= \frac{-i}{4(2\pi)^6} \iint \left(\pfrac{}{x^\mu} \pfrac{}{x_\mu} - m^2\right) \left(\pfrac{}{y^\mu} \pfrac{}{y_\mu} - m^2\right) \Delta_F(x-y) \delta(k_\nu k^\nu - m^2) \delta(l_\nu l^\nu - m^2) e^{ik_\xi x^\xi - il_\xi y^\xi} d^4x d^4y.
\end{aligned}
\end{equation}
From the first to the second line note that the operators are already ``time-ordered'', so one can equally well replace the canonical vacuum with the symmetric one. Then the result from (\ref{InOut}) can be used as in a time-ordered product this is all that remains because the in- resp. out-modes annihilate the opposite vacuum. Finally, the result can be written according to eq. (\ref{FeynmanProp}).

\end{document}